# Rapid diagnostics of reconfigurable intelligent surfaces using space-time-coding modulation

Yi Ning Zheng[1,†], Lei Zhang[1,†,*], Xiao Qing Chen[1], Marco Rossi[2], Giuseppe Castaldi[2], Shuo Liu[1], Tie Jun Cui[1,*], and Vincenzo Galdi[2,*]

[1] *Institute of Electromagnetic Space and State Key Laboratory of Millimeter Waves, Southeast University, Nanjing 210096, China*

[2] *Fields & Waves Lab, Department of Engineering, University of Sannio, I-82100 Benevento, Italy*

[†] *These authors contributed equally to this work.*

\* Corresponding authors: E-mail: njzhanglei@seu.edu.cn, vgaldi@unisannio.it, tjcui@seu.edu.cn.

## Abstract

Reconfigurable intelligent surfaces (RISs) have emerged as a key technology for shaping smart wireless environments in next-generation wireless communication systems. To support the large-scale deployment of RISs, a reliable and efficient diagnostic method is essential to ensure optimal performance. In this work, a robust and efficient approach for RIS diagnostics is proposed using a space-time coding strategy with orthogonal codes. The method encodes the reflected signals from individual RIS elements into distinct code channels, enabling the recovery of channel power at the receiving terminals for fault identification. Theoretical analysis shows that the normally functioning elements generate high power in their respective code channels, whereas the faulty elements exhibit significantly lower power. This distinction enables rapid and accurate diagnostics of elements' operational states through simple signal processing techniques. Simulation results validate the effectiveness of the proposed method, even under high fault ratios and varying reception angles. Proof-of-principle experiments on two RIS prototypes are conducted, implementing two coding strategies: direct and segmented. Experimental results in a realistic scenario confirm the reliability of the diagnostic method, demonstrating its potential for large-scale RIS deployment in future wireless communication systems and radar applications.

## 1. Introduction

Reconfigurable intelligent surfaces (RISs) are engineered structures composed of digital elements capable of manipulating electromagnetic (EM) waves in programmable manner to shape the propagation environment. Recently, RISs have garnered significant research interest and are recognized as a promising technology for next-generation wireless communications, due to their cost-effectiveness and ability to dynamically control signal transmission.[1-13] The hardware implementation of RISs largely relies on metasurfaces, the two-dimensional counterpart of metamaterials. Metamaterials are artificial EM structures composed of periodic or quasi-periodic subwavelength meta-atoms, allowing precise manipulations of EM waves. This capability enables the realization of advanced EM phenomena and applications including perfect lenses, negative refraction, illusion devices, and invisibility cloak.[14-18] Cui *et al.* [19] firstly introduced the concept of "digital coding" and "programmable" metasurfaces in 2014 by discretizing the responses of meta-atoms and incorporating switchable components such as diodes. Programmable metasurfaces (PMs), controlled by reconfigurable devices, offer real-time manipulations of EM waves while maintaining an intuitive design and a simple hardware architecture. Such features enable wide applications beyond the wireless communications, including vortex beam generation,[20] self-adaptive devices,[21-23] programmable holograms, [24,25] intelligent imaging,[26,27] and programmable artificial intelligence EM structures.[28] While early studies were focused on time-invariant coding patterns, *space-time-coding* (STC) digital metasurfaces have recently attracted increasing research interests by incorporating modulations across both spatial and temporal domains.[29,30] By dynamically varying digital states of the meta-atoms in time and space, the STC digital metasurfaces can simultaneously control the spatial and spectral distributions of EM waves, thus enabling a range of promising applications, including harmonic control,[30-34] scattering reduction,[30,35] reprogrammable non-reciprocity,[36] simplified-architecture wireless transmitters,[2,37] microwave sensing,[38-43] manipulation of all EM wave characteristics,[44, 45] and integrated sensing and communication systems.[46]

Despite all benefits and advances that RISs or PMs can potentially introduce to wireless communication technology, their practical deployment remains challenging. The performance of RISs can deteriorate, sometimes drastically, due to the factors such as dirt accumulation in outdoor environments or element malfunctions.[47-51] A similar challenge also affects the traditional antenna arrays, such as phased arrays or reflectarrays.[52-56] Therefore, a reliable RIS diagnostic method is essential for identifying faulty elements, allowing for efficient component replacement or compensation through software algorithms. A well-established approach to RIS diagnostics is focused on the signal level. [47-49,52] By leveraging the statistical characteristics of transmitted and received signals, along with the channel state information, researchers have successfully identified the operational states (normal or faulty) of individual elements, primarily using algorithms based on compressive sensing (CS). These diagnostic methods, often integrated with the fundamental communication processes such as channel estimation[47] and signal recovery[48], typically involve iterative algorithms and complicated signal-processing techniques, resulting in high computational demands. Moreover, most of these studies have been validated solely through theoretical analysis and simulations. Hardware-based methods diagnose RIS performance by measuring physical parameters such as far-field[53,55,56] or near-field[54] patterns. These approaches identify faulty elements by comparing measurement samples of the structures under test with those of fault-free ones. However, they rely on prior measurements of properly functioning structures and require accurate measurements at specific locations. Additionally, these methods often assume that faults are *sparse*, i.e., the number of faulty faults is significantly lower than the total number of elements, thereby posing limitations in practical applications. In these techniques, element states are typically considered as *static* throughout the diagnostic process. Recent studies have addressed the need for multiple measurements at different spatial locations by reconfiguring the states of individual elements.[50, 51] By sequentially switching the reflection phase of coding elements and comparing complex or amplitude-only received signals across different iterations, researchers have been able to identify faulty elements. While this approach allows the receiving antenna to remain fixed in one location, the reliance on iterative measurements slows down the diagnostic process. Additionally, the need for extended time-domain sampling or multiple measurements at different positions imposes practical limitations and introduces inconvenience for real-world applications.

Against this background, we propose here a STC modulation strategy for RIS diagnostics based on orthogonal codes. This method employs time-varying control signals to encode the reflected signals of different elements into independent channels. Faulty elements, which fail to modulate signals correctly, result in blank corresponding code channels, enabling effective fault identification. A detailed theoretical framework and mathematical model of the method are provided in the following section. Numerical simulations are conducted to analyze the impact of signal-to-noise ratio (SNR), fault count, and receiving angle, followed by experimental validation using two RIS prototypes. Unlike previous approaches, this method eliminates the need for complex algorithms or iterative measurements, offering advantages such as short testing time, simple implementation, low computational requirements, and high robustness. These characteristics make it particularly well-suited for practical application scenarios, such as the factory inspection and routine RIS testing.

## 2. Theory and modeling

**Figure 1** illustrates the fundamental concept underlying the proposed RIS diagnostic method, which utilizes a reflection-type PM undergoing spatio-temporal modulation. This is obtained by dynamically controlling the state of reconfigurable elements (e.g., diodes) via a field-programmable gate array (FPGA). As shown in the left part of the figure, the modulation of the RIS elements is described by a three-dimensional STC matrix based on orthogonal codes. When the RIS is illuminated by a monochromatic wave, the reflected signal is received and analyzed to determine the power of each code channel. Since faulty elements fail to modulate properly, the power of their corresponding code channels remains significantly low. By analyzing the power distribution across these channels, faulty elements can be accurately identified. The following section presents a detailed theoretical model of the RIS diagnostic procedure.

### 2.1 Coding strategy

As the RIS under test, we consider a 1-bit PM with $N$ independent coding elements. Here, an independent coding element refers to a group of meta-atoms that share the same control signal from the FPGA control module, such as meta-atoms within the same column of a column-controlled PM or a single meta-atom in a point-controlled PM. These coding elements are modulated by different Hadamard codes, which are widely used in the code-division multiple-access communication systems and signal processing.[57] The Hadamard codes are derived from rows of a Hadamard matrix $\mathbf{H}_K$, where $K$ represents the code order, and exhibit strong orthogonality, making them highly suitable for this application. A $K$-order Hadamard matrix is recursively defined as:

$$\mathbf{H}_K = \begin{bmatrix} \mathbf{H}_{K/2} & \mathbf{H}_{K/2} \\ \mathbf{H}_{K/2} & -\mathbf{H}_{K/2} \end{bmatrix}, \quad \mathbf{H}_1 = 1. \tag{1}$$

Equation (1) defines an orthogonal matrix whose rows represent mutually orthogonal Hadamard codes. It also demonstrates that Hadamard codes of length $K$, consisting solely of $\pm 1$ elements, correspond to binary phase states (0 or $\pi$) in complex notation.

This inherent phase discretization allows for direct implementation using 1-bit PMs, where the π-phase-shift capability matches the phase requirements of the codes.

It is important to note that not all Hadamard codes are suitable for this application. According to Equation (1), the first row of a Hadamard matrix consists entirely of +1, i.e., it does not introduce temporal modulation. As a result, elements modulated by this code would exhibit no measurable difference between normal and faulty conditions, except for a $\pi$ phase shift, which is impractical to detect. To address this limitation, codes from the first rows of Hadamard matrices are excluded, and the $n$th coding element is instead assigned the $(n+1)$th row of a Hadamard matrix. When all elements are modulated simultaneously, the minimum required order of Hadamard codes is given by

$$K_{\min} = 2^{\lceil \log_2(N+1) \rceil}, \tag{2}$$

where $\lceil \cdot \rceil$ denotes the ceiling function. Thus, for example, for a RIS with 15 elements, a 16th-order Hadamard code is needed. As $N$ increases, higher-order codes become necessary, leading to greater computational resource demands. A straightforward approach to reducing the required code order is to divide the RIS into smaller segments and diagnose each segment sequentially.

To implement Hadamard code modulation in the time domain, the control sequence $\mathbf{c}_n$ in the $n$th element of a PM must satisfy

$$c_n^l = \begin{cases} 0, & \mathbf{H}_K(n+1, l) = -1, \\ 1, & \mathbf{H}_K(n+1, l) = +1, \end{cases} \tag{3}$$

where $c_n^l$ denotes the bit symbol of $\mathbf{c}_n$ in the $l$th time slot. Here, the conventional mapping is intentionally reversed (typically, +1 maps to state 0 and -1 to state 1) to enhance synchronization, as discussed in Section 3. This modification does not affect the orthogonality of the coding scheme. Under the control sequence $\mathbf{c}_n$, the reflection coefficient of the $n$th element during the $l$th time slot is given by

$$\begin{aligned} \Gamma_n^l &= A_e \exp[j(\phi_0 + c_n^l \pi)] \\ &= -\mathbf{H}_K(n+1, l) A_e \exp(j\phi_0) \end{aligned} \tag{4}$$

where $A_e$ represents the reflection amplitude, $\phi_0$ denotes the reflection phase of state "0", and a time-harmonic dependence $\exp(j2\pi f_c t)$ is assumed. Given that the symbol duration in the modulation sequence is $\tau_0$, the time-varying reflection coefficient $\Gamma_n(t)$ of an element controlled by the sequence in Equation (3) can be expressed as:

$$\begin{aligned}\Gamma_n(t) &= \sum_{l=1}^{K} \Gamma_n^l G^l(t) \\ &= -A_e \exp(j\phi_0) \sum_{l=1}^{K} \mathbf{H}_K(n+1, l) G^l(t),\end{aligned} \qquad (5)$$

where $G(l, t)$ represents the pulse gate function, defined as

$$G^l(t) = \begin{cases} 1, & (l-1)\tau_0 \leq t < l\tau_0, \\ 0, & \text{otherwise.} \end{cases} \qquad (6)$$

**2.2 Signal model**

Assuming a monochromatic plane wave at frequency $f_c$ impinging on the RIS from direction $(\theta_i, \varphi_i)$, the received signal at a receiver positioned at $(\theta_o, \varphi_o)$ is given by

$$r(t) = s(t) \sum_{n=1}^{N} \Gamma_n(t) E_n^i a_n(\theta_i, \varphi_i) E_n^o a_n(\theta_o, \varphi_o) + n(t), \qquad (7)$$

where $E_n^i$ and $E_n^o$ represent the angular response of the $n$th element in the incident and outgoing directions, respectively. The incident signal is expressed as $s(t) = \exp(j2\pi f_c t)$, $n(t)$ is an additive Gaussian white noise, and $a_n(\theta, \varphi)$ is the steering factor of the $n$th element, given by

$$a_n(\theta, \varphi) = \exp\left[j\frac{2\pi}{\lambda_c}(d_{nx} \sin\theta \cos\varphi + d_{ny} \sin\theta \sin\varphi)\right], \qquad (8)$$

where $d_{nx}$ and $d_{ny}$ denote the distances of the $n$th coding element from the first element along $x$ and $y$ directions, respectively. Equation (7) can be discretized with a time interval $\tau_0$ as

$$r_l = \sum_{n=1}^{N} \Gamma_n^l E_n^i E_n^o a_n(\theta_i, \varphi_i) a_n(\theta_o, \varphi_o) + n_l. \qquad (9)$$

For a normally functioning element, $\Gamma_n(t)$ follows a periodic $K$-order Hadamard code sequence, as described in Equation (5). Conversely, for a faulty element, the reflection coefficient remains time-invariant, i.e., $\Gamma_n(t) = A_e \exp j\phi_0$, which implies a direct structural relationship with $\mathbf{H}_K(1,\cdot)$, composed entirely of +1, through the decomposition

$$\Gamma_n(t) = A_e \exp(j\phi_0) \sum_{l=1}^{K} \mathbf{H}_K(1,l) G^l(t). \tag{10}$$

Equation (10) reveals that the reflected signal components from faulty elements remain orthogonal to those from normal elements due to the inherent orthogonality of $\mathbf{H}_K$. This intrinsic energy decoupling ensures that faulty elements introduce no interference in the code channel energy of normal elements, thus enabling precise and unambiguous RIS diagnostics.

**2.3 RIS diagnostics**

At the receiving terminal, the signal power of the $m$th coding channel, $\hat{P}_m$, is recovered as

$$\hat{P}_m = \left| \frac{1}{K} \sum_{l=1}^{K} \mathbf{H}_K(m+1,l) r_l \right|^2. \tag{11}$$

Substituting Equation (9) into Equation (11) and interchanging the summation order, we obtain

$$\hat{P}_m = \left| \sum_{n=1}^{N} E_n^i E_n^o a_n(\theta_i, \varphi_i) a_n(\theta_o, \varphi_o) \frac{1}{K} \sum_{l=1}^{K} \mathbf{H}_K(m+1,l) \Gamma_n^l + n_l \right|^2. \tag{12}$$

Analyzing Equation (12), two cases arise:

- **Case 1**: The $m$th element is normally functioning, i.e., $\Gamma_m^l = -\mathbf{H}_K(m+1,l) A_e \exp(j\phi_0)$.

Using Equation (10) and the orthogonality of $\mathbf{H}_K$,

$$\frac{1}{K} \sum_{l=1}^{K} \mathbf{H}_K(m+1,l) \Gamma_n^l = \begin{cases} 0, & n \neq m, \\ -A_e \exp(j\phi_0), & n = m. \end{cases} \tag{13}$$

Thus, the recovered code channel power is

$$\hat{P}_n = \left| E_n^i E_n^o a_n(\theta_i, \varphi_i) a_n(\theta_o, \varphi_o) A_e \right|^2 + P_n, \tag{14}$$

where $P_n$ denotes the noise power.

- **Case 2**: The *m*th element is faulty, i.e., $\Gamma_m^l = \mathbf{H}_K(1, l) \cdot A_e \exp(j\phi_0)$.

From Equation (10) and the orthogonality of $\mathbf{H}_K$, we obtain

$$\frac{1}{K}\sum_{l=1}^{K} \mathbf{H}_K(m+1, l)\Gamma_n^l = 0. \tag{15}$$

which leads to

$$\hat{P}_m = P_n. \tag{16}$$

By comparing Equation (14) and (16), it can be concluded that in a high SNR scenario, an appropriate threshold $\Lambda$ can be set to determine the state of the *m*th element. If $\hat{P}_m > \Lambda$, the element is diagnosed as normally functioning; if $\hat{P}_m < \Lambda$, it is identified as faulty. Importantly, the proposed method is robust to variations in the incident angle $\theta_i$ and the receiving angle $\theta_o$. While these angles affect the *absolute* intensity $|E_n^i E_n^o a_n(\theta_i, \varphi_i) a_n(\theta_o, \varphi_o) A_e|^2$, the diagnostic process relies on the *relative* intensity difference between the normal and faulty elements, ensuring consistent and reliable fault detection.

In summary, the proposed diagnostic procedure consists of the following steps:

1. Modulate each coding elements with a unique Hadamard code as defined in Equations (1)-(5).
2. Illuminate the RIS under test with a monochromatic plane wave and receive the reflected signal.
3. Compute the power of each code channel $\hat{P}_m$ using Equation (11).
4. Diagnose the *m*th element based on the threshold $\Lambda$

$$\begin{cases} \text{Normal, if } \hat{P}_m > \Lambda, \\ \text{Faulty, if } \hat{P}_m < \Lambda. \end{cases}$$

This diagnostic method does not rely on the assumption of sparsity commonly used in CS-based approaches, making it suitable for RIS systems with a high fault ratio.[47-49, 52-55]

**Table 1.** Design parameters of RIS prototypes for fault diagnostics.

|  | Prototype 1 | Prototype 2 |
|---|---|---|
| **Testing frequency** | 9.8 GHz | 10 GHz |
| **Array dimensions** | $16 \times 16$ | $8 \times 8$ |
| **Meta-atom size** | 15mm × 15mm | 15mm × 15mm |
| **Phase resolution** | 1 bit | 1 bit |

## 3. Results and Discussions

### 3.1 Simulation results

To validate the proposed diagnostic method, we first conduct a numerical simulation using a $16 \times 16$ column-controlled metasurface as the RIS under test. This metasurface operates at $9.8 \text{ GHz}$, and consists of $15\text{mm} \times 15\text{mm}$ meta-atoms, matching the configuration of Prototype 1 in **Table 1**. Following Equation (2), we set the code order to 32, with each code symbol lasting $\tau_0 = 1 \text{ µs}$, corresponding to a modulation frequency of $f_0 = 62.5 \text{ kHz}$. The digital sampling frequency is $f_s = 8 \text{ MHz}$. We set the incident angle to $\theta_i = 0°$ and position the receiver at $\theta_o = 20°$ relative to the RIS normal. The SNR of the incident signal is set at 30dB.

Building on the strategy described in the previous section, we implement space-time modulation of the RIS using the STC matrix shown in **Figure 2**a. This matrix is derived from a 32nd-order Hadamard matrix, excluding the first row. To simulate faulty elements, we set random rows in the STC matrix to all zeros. By modifying seven rows in this way, Figures 2b and 2c illustrate the amplitude and phase of the time-domain received signal over six modulation periods, while Figure 2d shows the operational states of each element. The received signal is processed using Equation (11) to extract the power of each coding channel. To reduce noise effect, we compute the mean power over the six modulation periods. Figure 2e displays the recovered power across different coding channels, which clearly separates into two distinct groups. Since the diagnostics is based on relative intensity rather than absolute values, the power is normalized with respect to the maximum recovered power. Comparing Figure 2e and 2d, we observe that the normalized power of channels corresponding to normal elements remains close to 0dB, whereas faulty element channels exhibit significantly lower power. By setting an appropriate threshold, we can easily identify faulty elements based on their low recovered power, which is in perfect agreement with the actual fault distribution and confirms the effectiveness of the proposed method. Furthermore, the proposed method is scalable to large-scale point-controlled RISs, with corresponding results provided in the Supporting Information (see **Figure S1**).

We further conduct simulations to evaluate the impact of SNR and the number of faulty elements, $N_F$, on the diagnostics error rate (ER). This latter is defined as the average ratio of incorrectly diagnosed elements to the total number of elements over 3000 simulation runs. **Figure 3**a shows that for a fixed $N_F$, the ER decreases as SNR increases. When SNR exceeds 15dB, the ER drops below $10^{-4}$, regardless of $N_F$. On the other hand, as the number of faulty elements increases, the ER remains nearly constant or even slightly decreases. These results confirm that the proposed method can reliably diagnose RIS faults, even in scenarios with a high fault ratio.

Additionally, we simulate the ER across different receiving angles $\theta_o$ while keeping the incident angle fixed at $\theta_i = 0$. As shown in Figure 3b, although smaller receiving angles lead to better ER performance in low-SNR conditions, the ER remains stable within the range $-30° \leq \theta_o \leq 30°$ when SNR > 10dB, further supporting the previously inferred robustness to angle variations. Moreover, for SNR > 20dB, the ER remains below $10^{-4}$ across the entire range $-60° \leq \theta_o \leq 60°$.

These results confirm the feasibility and robustness of the proposed method. In practical applications, achieving an SNR above 15dB can be easily accomplished by increasing the transmission power. Consequently, the RIS diagnostic method remains effective across various SNR conditions, multiple receiving angles, and different fault densities. Moreover, it requires only a single measurement to collect the data and enables rapid fault detection with minimal processing.

## 3.2 Experimental validation

To further validate the proposed RIS diagnostic method, we implement it on two different RIS prototypes. **Figure 4**a illustrates the architecture of the diagnostic procedure. A pair of standard horn antennas serve as transmitting and receiving units. One signal generator produces the monochromatic transmission signal, while another supplies the local-oscillator signal for down-conversion. The received signal is acquired by a universal software radio peripheral (USRP) and processed by a host computer. Figure 4b shows the measurement setup in a real-world testing environment.

The detailed design parameters of the two RIS prototypes used in the experiments are listed in Table 1. Prototype 1 is a $16 \times 16$ column-controlled PM with a total aperture of $240\text{mm} \times 240\text{mm}$, whereas Prototype 2 features an $8 \times 8$ column-controlled PM with a total aperture of $120\text{mm} \times 120\text{mm}$. Both RISs operate with a 1-bit phase shift achieved by toggling the positive-intrinsic-negative (PIN) diodes soldered on each element between the "ON" and "OFF" states. Close-up views of both RISs are shown in Figures 4c and 4d. Further details on the designs and the impact of the unavoidable response imperfections are provided in the Supporting Information (see **Figure S2** and **Figure S3**).

Different coding schemes are applied to the two prototypes to account for variations in array scales and mutual coupling. For Prototype 2, which has a relatively small scale requiring only 16th-order codes, we use a direct coding scheme similar to the simulation setup. The STC matrix is derived from 16th-order Hadamard codes. For Prototype 1, we implement a segmented coding scheme to reduce the code order and mitigate the impact of mutual coupling between adjacent elements. **Figure 5**a illustrates the structure of a control signal frame divided into eight segments, including two padding bit segments and six segments of STC signals that alternately modulate odd and even columns. Since only half of the columns are modulated within a segment, the same codes can be reused across different segments, effectively reducing the code order from 32 to 16. Additionally, as adjacent elements remain static when one element is modulated, interference caused by mutual coupling is minimized. Two 16-bit padding segments with "0" symbols are inserted for synchronization. Figure 5b shows the detailed STC matrix within a frame of 128 bits, where 16th-order Hadamard codes are employed. The modulation symbol duration is set to $\tau_0 = 1\mu s$, corresponding to a switching frequency $f_{\text{switch}} = 1$ MHz. Under this control signal, the amplitude and phase of the received signal measured within a frame duration are shown in Figures 5c and 5d, respectively, with a sampling frequency of $f_s = 10$ MHz. It is worth noting that the phase hop at the end of the padding bits, caused by the reverse symbol mapping, simplifies the synchronization process. Additionally, similar to the simulation setup, we set the control signals to zero to simulate faulty elements in both prototypes.

**Figure 6** presents the experimental results for both RIS prototypes under various fault conditions, with different number of faulty elements. For Prototype 1, a total of twelve cases were measured, with Case 1 representing the fault-free condition. The actual operational states for each case are illustrated in Figure 6e. Figure 6a illustrates the received power pertaining to different columns, revealing that the recovered power drops sharply to below $-20$ dB when an element is faulty. By setting an appropriate threshold, faulty elements can be clearly distinguished. In this experimental configuration, a noticeable power imbalance of up to 5dB is observed between the central and edge columns. This imbalance is attributable to edge effects and the relatively large aperture, which prevents the incident signal from being perfectly approximated by a plane wave. This issue become more significant as the metasurface scale increases. Since the substantial power contrast of more than 20 dB between normal and faulty states significantly exceeds the 5 dB fluctuations, normalization using reference power from Case 1, where all elements operate normally, is not strictly necessary. However, we normalize the recovered power to enhance visual differentiation in result presentation. As shown in Figure 6c, this normalization significantly stabilizes the power distribution, reducing the imbalance between columns to 2dB. Comparing the normalized power in Figure 6a and 6c with the actual operational states in Figure 6e, we observe a clear distinction between normal and faulty elements, with a power difference up to 20dB.

For Prototype 2, nine different cases are measured, as shown in Figure 6f. Figures 6b and 6d present the received and normalized power for each case, demonstrating a power difference up to 10dB between normal and faulty elements. This confirms that the proposed diagnostic method can accurately determine the operational states of RIS elements by applying a proper threshold. Notably, even when the fault ratio exceeded 50% (in Case 11 and 12 for Prototype 1, and Case 8 and 9 for Prototype 2) where the sparsity assumption of traditional CS-based methods no longer holds,[43-45,48-51] our approach continues to deliver accurate fault diagnostics. This result highlights the superiority of the proposed method in high-fault scenarios, in agreement with theoretical predictions. The impact of the number of modulation periods on diagnostics accuracy is explored in the Supporting Information (see **Figure S4**).

Overall, the experimental results confirm the feasibility of the proposed diagnostic method for different RISs in a realistic setting. For small-scale or those with low-mutual-coupling, the direct coding scheme is effective, whereas the segmented coding scheme is better suited for large-scale RISs or those with higher mutual coupling. Moreover, the method does not require an anechoic chamber or iterative measurements, making it well-suited for practical applications such as routine maintenance or factory inspection. It is important to note that while this study focuses on 1-bit configurations with 180° phase differences, Hadamard codes can be adapted to accommodate arbitrary phase shifts.[58] Hence, the proposed approach can, in principle, be extended to multi-bit configurations.

## 4. Conclusions

We have put forward a RIS diagnostic method based on STC modulation strategy, where each element is assigned a unique orthogonal Hadamard code. Normally functioning elements temporally modulate the incident signal using independent orthogonal codes, while faulty elements reflect the signal without modulation. At the receiving terminal, the power of each coding channel is recovered. Since the channel power corresponding to faulty elements is significantly lower, an appropriate threshold enables accurate diagnostics. To establish the effectiveness of the proposed method, we developed a mathematical model and conducted a theoretical feasibility analysis, followed by numerical simulations. Additional simulations investigating the impact of SNR, the number of faulty elements, and the receiving angle on the diagnostics ER confirmed that the method reliably identifies faults even in high-fault scenarios.

Further validation was performed through experiments on two RIS prototypes, using two different coding schemes: direct coding scheme for smaller RISs with low mutual coupling, and segmented coding scheme for larger RISs with higher mutual coupling. Both simulations and experimental results demonstrate that the proposed method offers a simple and intuitive measurement process, is highly practical for RISs with a high fault ratio, and remains robust across different receiving angles.

These advantages make the method well-suited for applications requiring fast and accurate diagnosis, such as factory inspections for detecting insufficient soldering or broken components, as well as preliminary fault localization in routine maintenance of outdoor RIS deployment. As the RIS technology continues to expand in the next-generation wireless communication systems, this diagnostic method will play a crucial role in ensuring their reliability and performance.

## 5. Experimental Section

*Modeling*: The numerical implementation of the theoretical analysis [Equation (1)-(16)] was carried out using MATLAB, with the far-field scattering pattern of the basic element approximated as $E(\theta, \varphi) = \cos\theta$.

*Measurements*: The experimental setup for method verification is shown in Figure 4a, with detailed designs and measured reflection coefficients provided in the Supporting Information (see Figure S2). A pair of X-band standard horn antennas serve as the transmitter and receiver. An Agilent E8257D signal generator produces the monochromatic transmitting signal, while the received signal is acquired by a universal software radio peripheral (USRP, National Instrument NI-USRP 2943R) and subsequently processed on a host computer.

## Acknowledgment

Y.N.Z. and L.Z. contributed equally to this work. This work was supported by the National Key Research and Development Program of China (2023YFB3811504), the National Natural Science Foundation of China (62288101, 62101123, U22A2001, and 62201136), the Jiangsu Province Frontier Leading Technology Basic Research Project (BK20212002), the 111 Project (111-2-05), the National Postdoctoral Program for Innovative Talents (BX2021062), the Young Elite Scientists Sponsorship Program by CAST (2020QNRC001), the China Postdoctoral Science Foundation (2020M680062), and the Jiangsu Planned Projects for Postdoctoral Research Funds (2021K058A). The work of G. C. and V. G. was partially supported by the European Union under the Italian National Recovery and Resilience Plan (NRRP) of NextGenerationEU, partnership on "Telecommunications of the Future" (PE00000001 - program "RESTART").

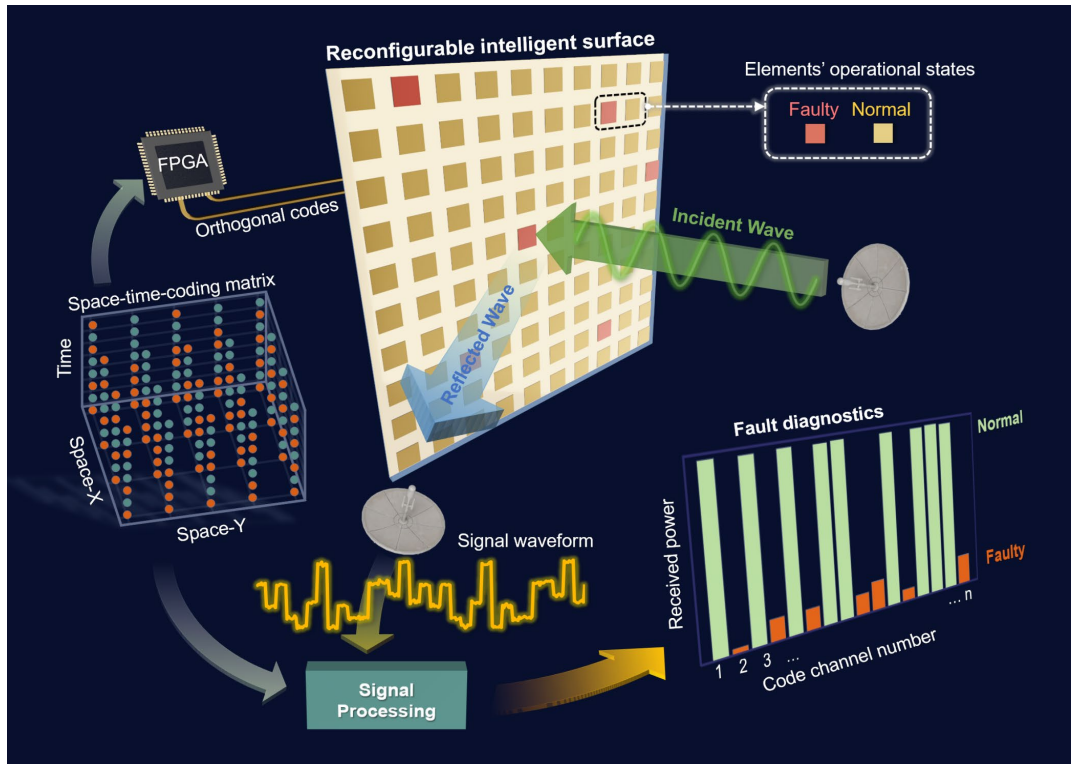

**Figure 1.** Conceptual illustration of RIS diagnostics based on STC modulation strategy. By assigning different orthogonal Hadamard codes to individual coding elements, faulty elements can be rapidly identified based on the recovered power of each code channel in the received signal.

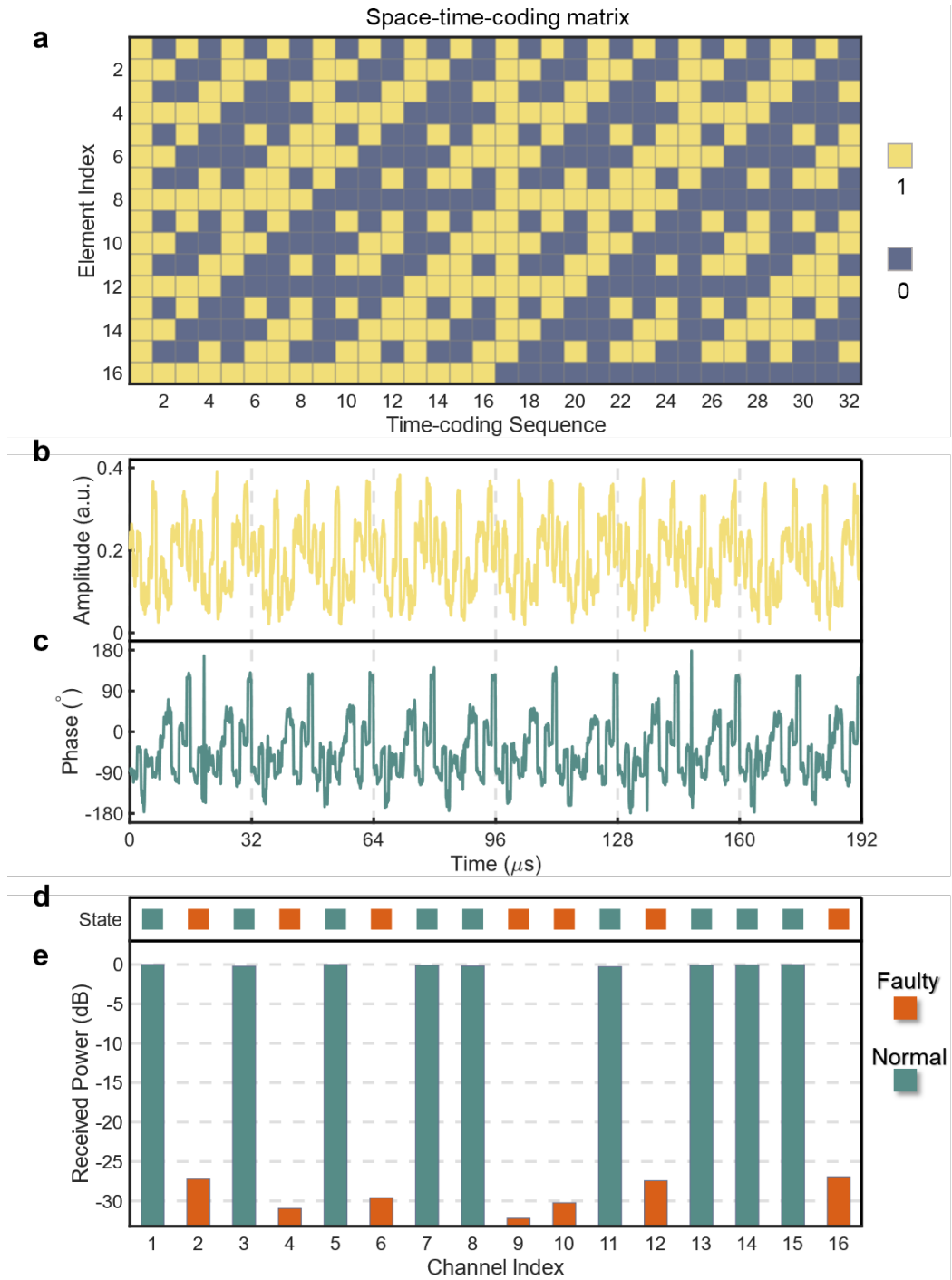

**Figure 2.** Simulation results for signal analysis and processing. Each column is modulated by a 32nd-order Hadamard code. a) STC matrix constructed from Hadamard codes. b,c) Amplitude and phase, respectively, of received time-domain signal over six modulation periods. d) Actual operational states of each column. e) Recovered power of different code channels for fault diagnostics.

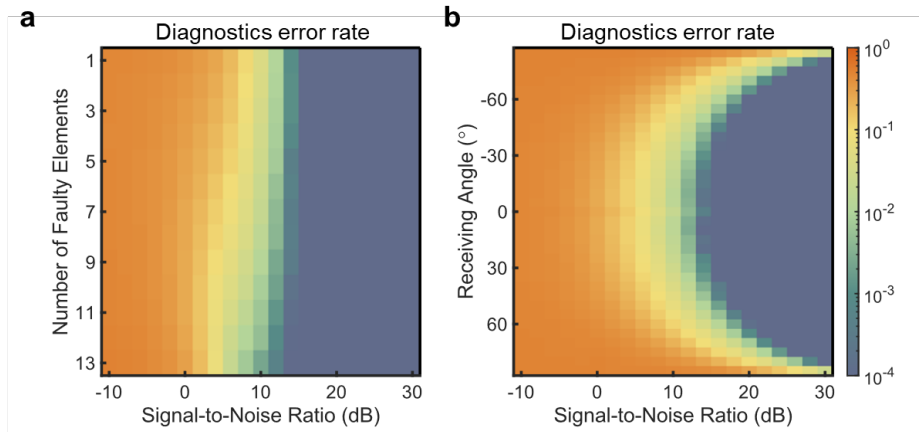

**Figure 3.** a,b) Simulated ER of the diagnostics under varying SNR conditions as a function of the number of faulty elements and the receiving angle, respectively.

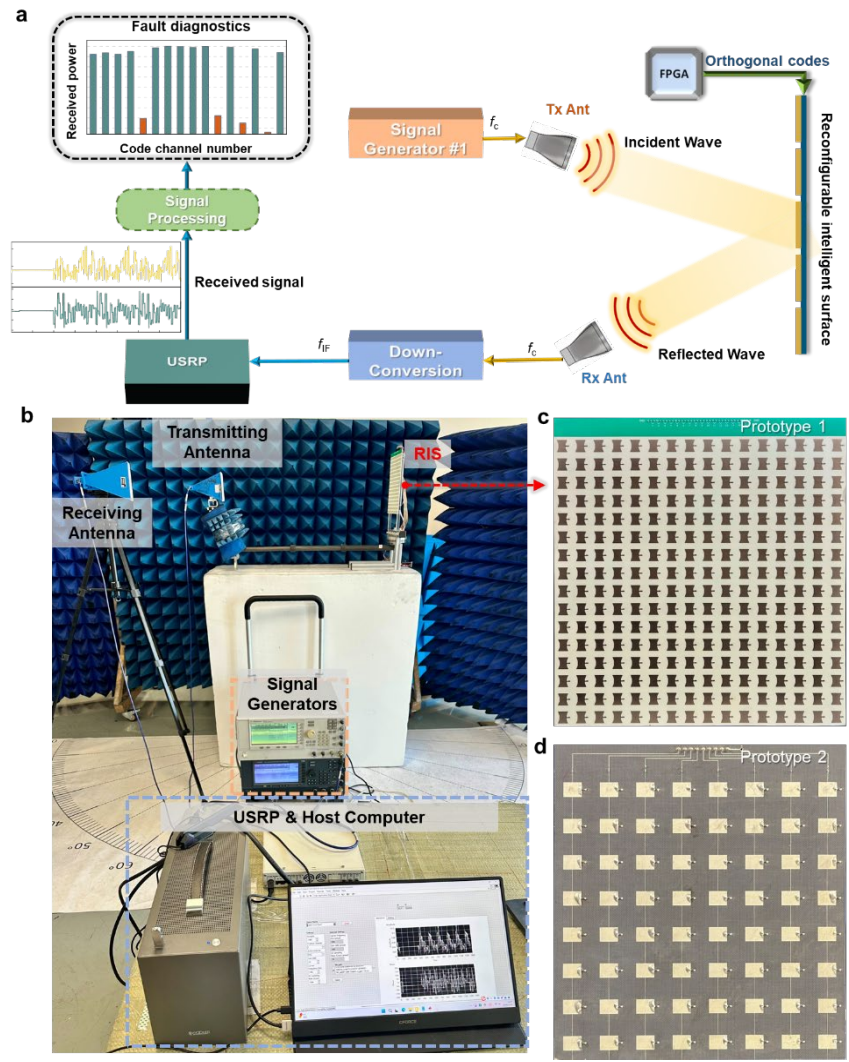

**Figure 4.** Measurement setup and RIS prototypes. a) Architecture of the proposed RIS diagnostic system. b) Photograph of measurement setup. c,d) Close-up views of RIS Prototypes 1 and 2, respectively.

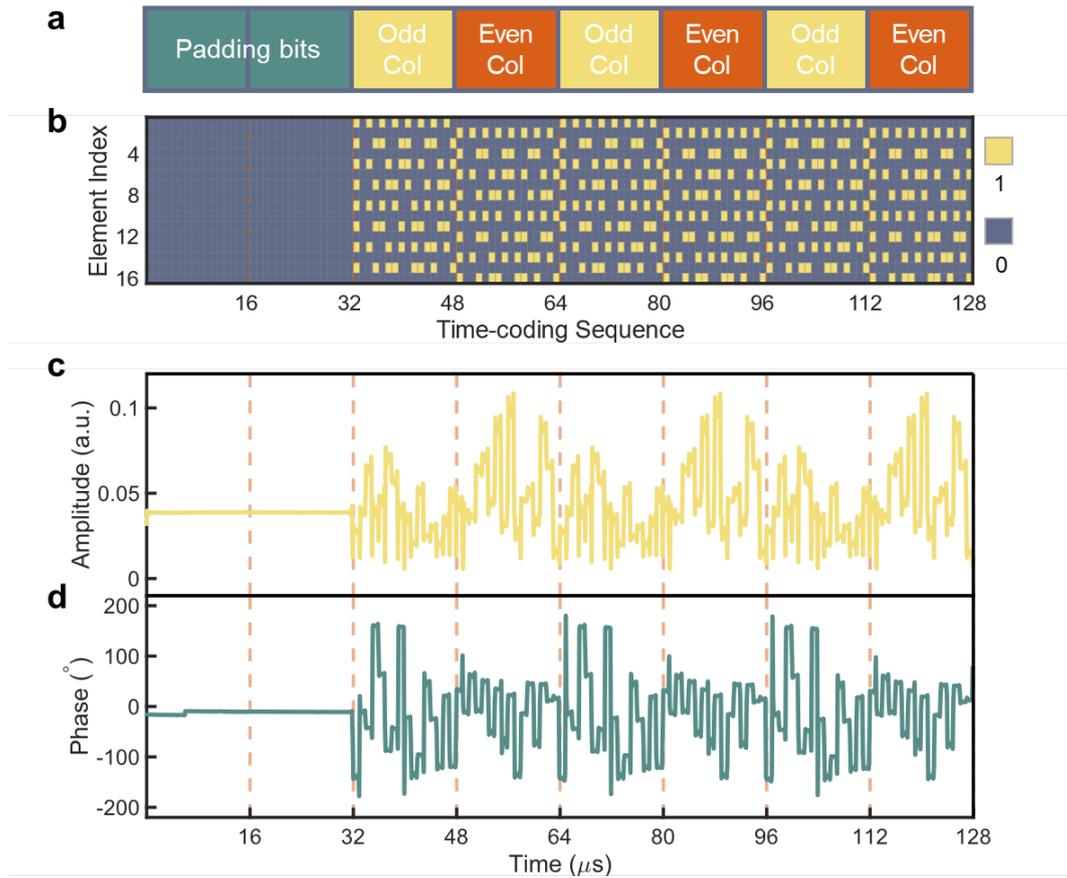

**Figure 5.** STC configuration in the experiment for Prototype 1. a) Modulation signal frame structure. b) Detailed STC matrix within a frame. c,d) Amplitude and phase, respectively, of received signal over a frame duration.

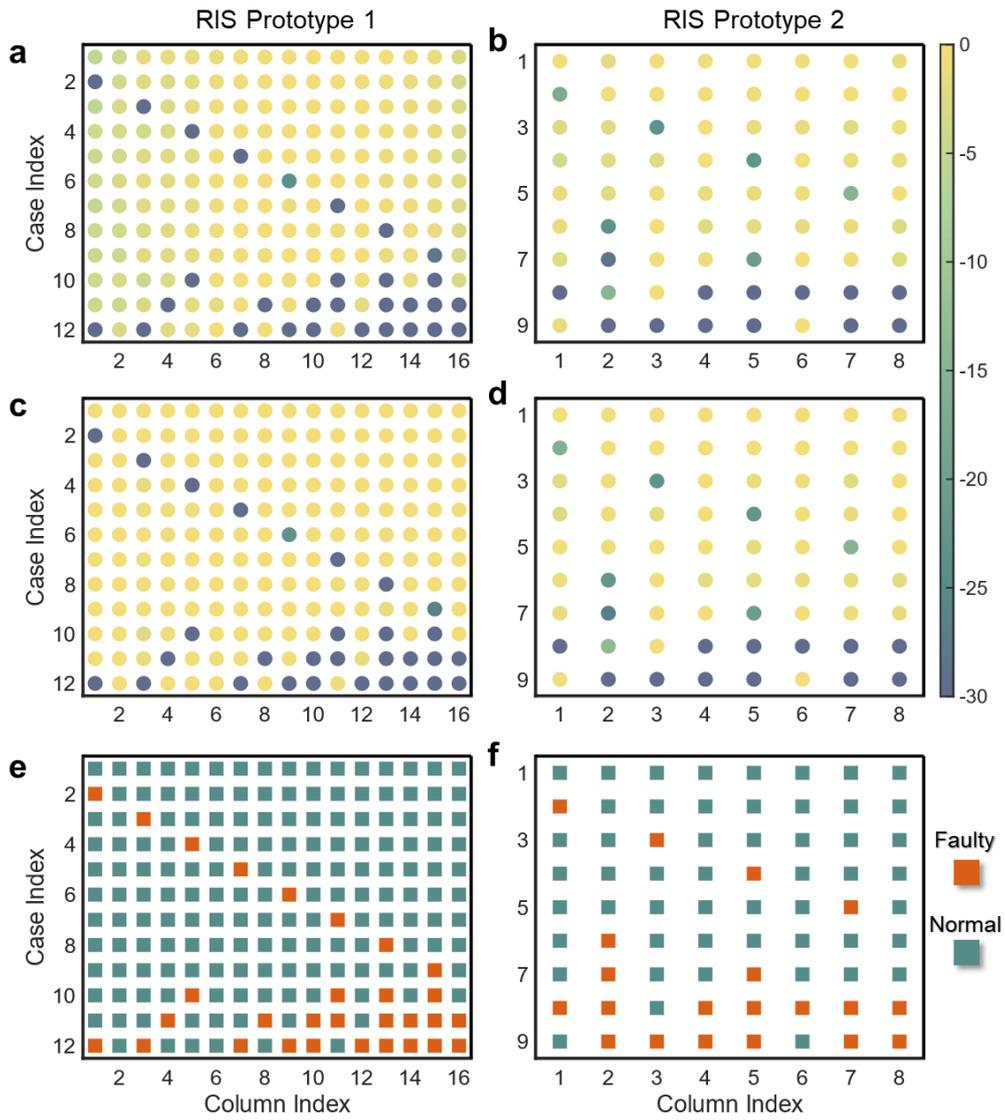

**Figure 6.** Experimental results of RIS diagnostics across different test cases. a,c,e) Recovered power, normalized power, and actual operations states, respectively, for Prototype 1 across twelve cases. b,d,f) Recovered power, normalized power, and actual operations states, respectively, for Prototype 2 across nine cases.

# Supplementary Information

## 1. Simulation validation of fault diagnostics in large-scale point-controlled RIS with high fault ratios

We further simulate the proposed method on large-scale point-controlled RISs, specifically, a $15 \times 15$ array (PC-RIS 1) and a $25 \times 25$ array (PC-RIS 2), each with approximately 30% randomly distributed faulty elements. The simulation parameters remain consistent with those in the main text. For both RISs, the direct coding scheme is applied, utilizing STC matrices based on 256th-order and 1024th-order Hadamard codes, respectively. Received signals over 20 modulation periods are collected and analyzed. **Figure S1**a and S1b show the average received power for each coding element's channel in PC-RIS 1 and PC-RIS 2, respectively, while Figures S1c and S1d present their actual operational states. The results indicate that faulty elements are reliably identified, as their recovered power falls below -15dB, confirming the effectiveness of the proposed diagnostic method even for large-scale RISs with high fault ratios.

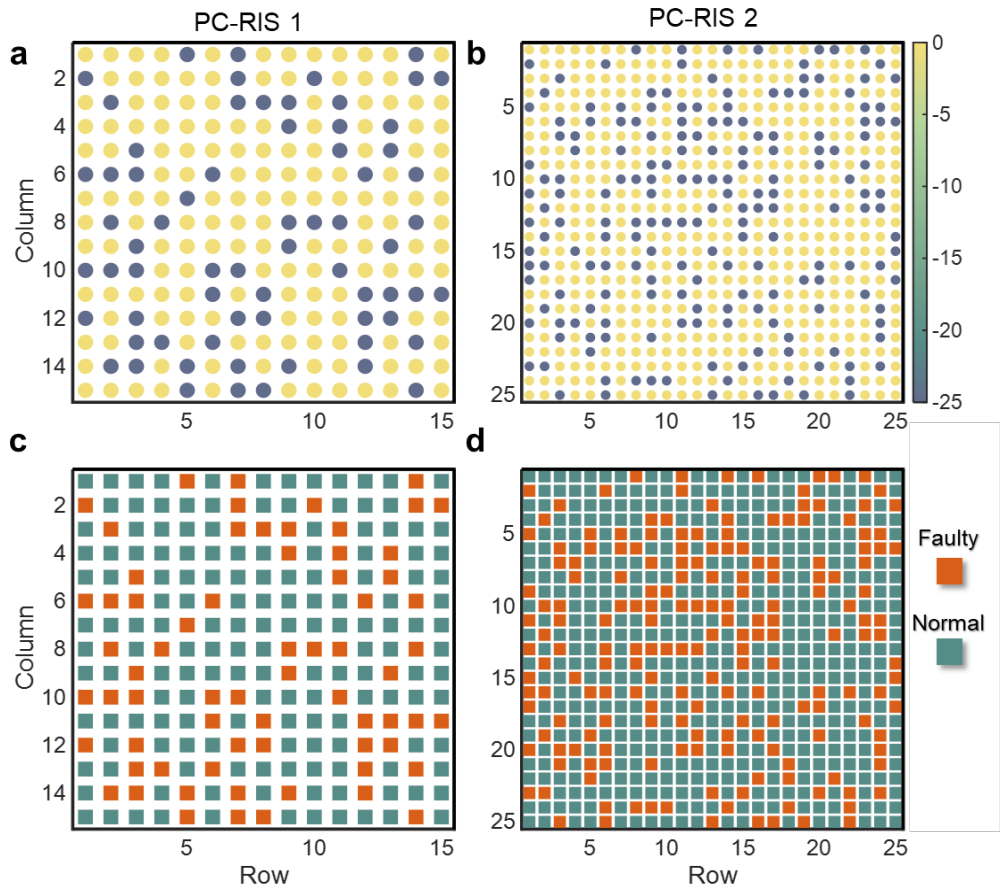

**Figure S1.** Simulation results for large-scale point-control RISs. a,b) Simulated recovered power of each coding element's channel in PC-RIS 1 PC-RIS 2, respectively. c,d) Corresponding actual operational states.

## 2. Detailed designs and element responses of RIS prototypes

**Figure S2**a illustrates the element design of RIS Prototype 1, which features a three-metal-layer structure sandwiched between two Rogers RO4003C dielectric layers and one FR4 prepreg layer. The programmable meta-atom consists of a PIN diode (MACOM MADP-000907-14020X) connecting the bow-tie patch to a grounded rectangular patch on the top layer, with bias voltage supplied through bottom-layer lines connected by via. This configuration enables 180° phase difference between diode's "ON" and "OFF" states. Numerical simulations conducted via CST Microwave Studio (https://www.3ds.com/ products/simulia/cst-studio-suite) (Figure S2c) show a 1-bit

phase response near 10 GHz with less than 1 dB amplitude loss. This is consistent with experimental results (Figure S2e), which confirm a 180° phase shift at 9.8 GHz.

Figure S2b illustrates the element design of RIS Prototype 2, which employs a two-metal-layer structure using an F4B dielectric. The top-layer diode (same as the previous design) connects the central rectangular patch to grounded vias, with bias voltage supplied directly through top-layer lines. Both CST simulations (Figure S2d) and experimental measurements (Figure S2f) confirm a 180° phase switching at 10GHz and 9.8 GHz, respectively, with an amplitude loss of approximately 2 dB in experiments.

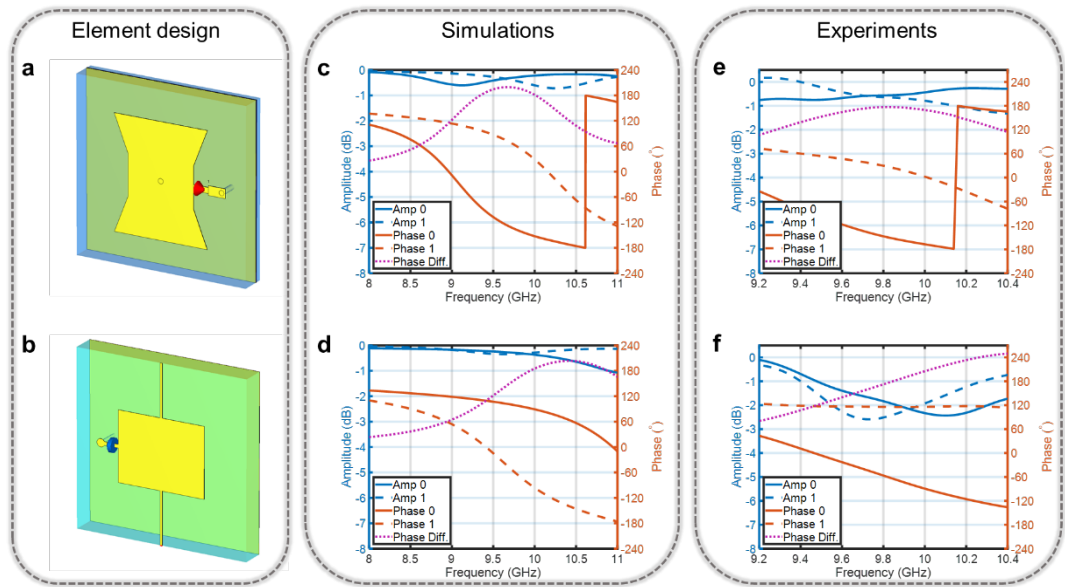

**Figure S2**. Designs and response characteristics of RIS prototypes. a,b) Element design of Prototype 1 and 2, respectively. c,d) Corresponding simulated responses. e,f) Corresponding measured responses.

## 3. Analysis of 1-bit phase deviations and reflection amplitude imbalance

Figures S2c-f reveal that within the operational frequency band of the RIS elements, the phase difference between states "0" and "1" deviates from the ideal 180°, accompanied by an imbalance in reflection amplitude. Additionally, manufacturing

tolerances cause frequency shifts between the simulated and measured operational frequencies. These non-ideal characteristics introduce imperfection in orthogonal code modulation during RIS diagnostics. To evaluate the impact of these effects, we simulate diagnostics outcomes under varying phase deviations (160°-200°) and amplitude imbalances, as shown in **Figure S3**. Figures S3a-c indicate that while phase deviations introduce energy fluctuations across code channels, the distinction between normal and faulty elements remains sufficiently clear for accurate diagnostics. Similarly, Figures S3d-f show that the diagnostic performance remains stable despite amplitude imbalances. These results confirm that the method retains high accuracy despite variations in element response, demonstrating its robustness.

Notably, at the testing frequency of 10 GHz (see Figure 6 in the main text), RIS Prototype 2 exhibits a phase difference of approximately 200° rather than 180° at its actual operational frequency of 9.8 GHz. Despite this deviation, the method successfully detects faults, further validating its tolerance to non-ideal element responses.

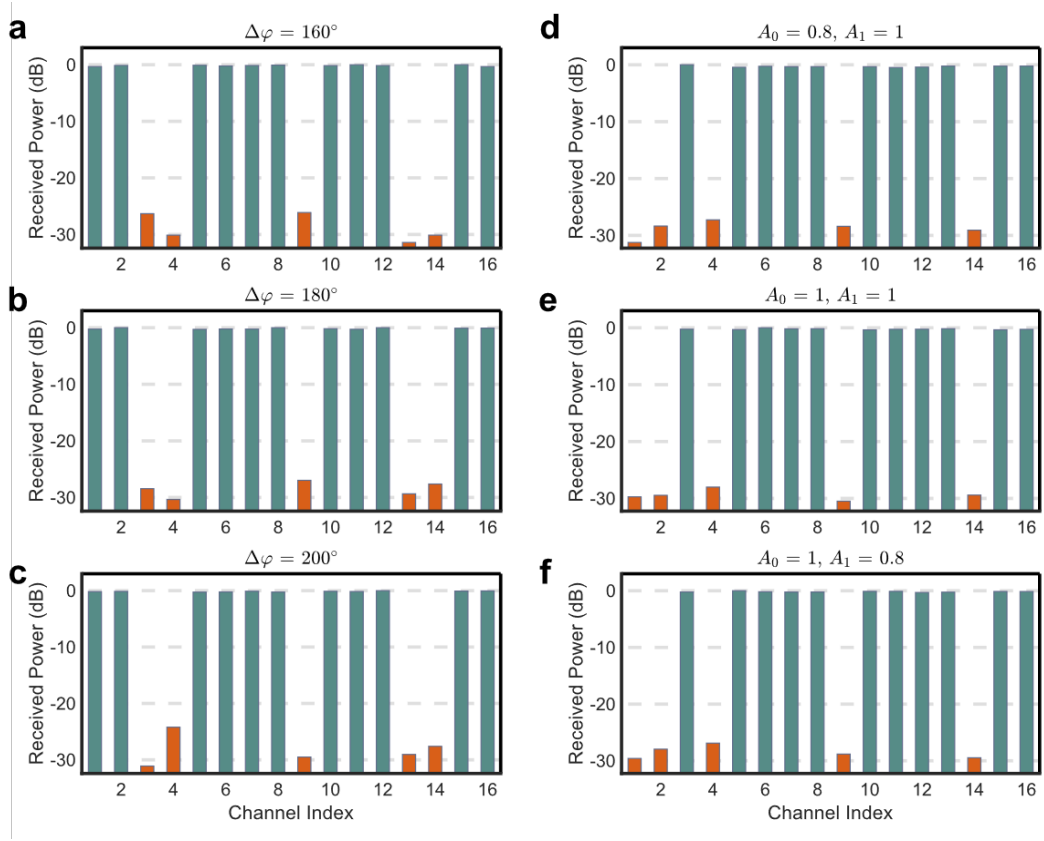

**Figure S3.** Simulation analysis of the impact of element response imperfections. a,b,c) Diagnostics results under varying phase differences. d,e,f) Diagnostics results under different reflection amplitude imbalances.

## 4. Impact of modulation periods on diagnostics performance

In the main text, the recovered power of both RIS prototypes in all test cases is calculated based on received signals within a single frame duration, which contains only six segments of valid modulated signal periods. To assess the impact of the number of modulation periods on diagnostics accuracy, numerical simulations are conducted with 1, 6, 12, and 200 periods, as shown in **Figure S4**. Comparing Figures S4a-d, it is evident that code channel power stabilizes with as the number of modulation periods increases: the power pertaining to normal elements converges toward 0 dB, while the power of faulty elements approaches noise energy levels (-30 dB in this simulation). However, even with a minimal number of periods (Figures S4a and S4b), the power difference between normal and faulty elements remains substantial. Additionally, the improvement between Figure S4c and S4d is marginal, indicating diminishing returns with further increases in modulation periods.

Including more modulation periods in a frame can help suppress noise effects, enhancing the consistency and reliability of the results. However, under high signal-to-noise ratio (SNR) conditions, a limited number of modulation periods is sufficient to achieve accurate diagnostics.

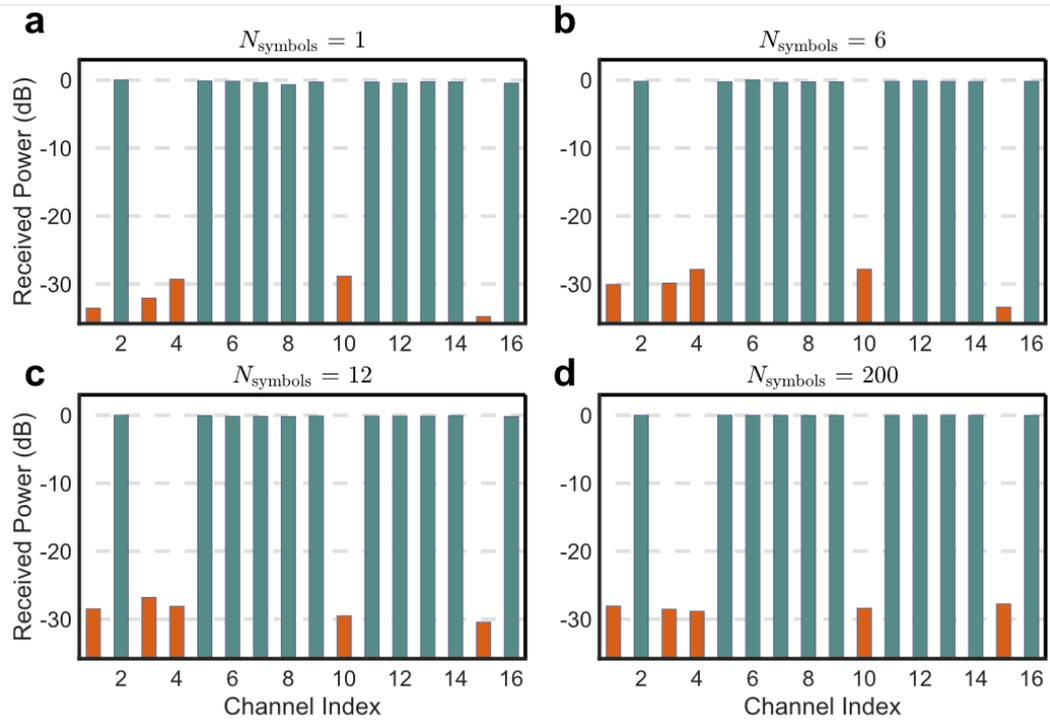

**Figure S4.** Simulation results of diagnostics with varying numbers of modulation periods per frame. a,b,c,d) 1, 6, 12, and 200 periods, respectively.